\documentclass[preprint,aps,prd,superscriptaddress,nofootinbib,tightenlines]{revtex4}

\usepackage{amsfonts}
\usepackage{multirow}
\usepackage{mathrsfs}
\usepackage{graphicx}
\usepackage{amsmath}
\usepackage{amssymb}
\usepackage{bm}
\usepackage{bbm}
\usepackage{color}
\usepackage{array}
\usepackage{diagbox}
\usepackage{float}

\def\OMIT#1{}

\def\hlinew#1{%
  \noalign{\ifnum0=`}\fi\hrule \@height #1 \futurelet
   \reserved@a\@xhline}
\usepackage{array}
\newcommand{\PreserveBackslash}[1]{\let\temp=\\#1\let\\=\temp}
\newcolumntype{C}[1]{>{\PreserveBackslash\centering}p{#1}}
\newcolumntype{R}[1]{>{\PreserveBackslash\raggedleft}p{#1}}
\newcolumntype{L}[1]{>{\PreserveBackslash\raggedright}p{#1}}

\newcommand{\nn}{\nonumber}

\newcommand{\beq}{\begin{equation}}
\newcommand{\eeq}{\end{equation}}
\newcommand{\bqa}{\begin{eqnarray}}
\newcommand{\eqa}{\end{eqnarray}}

\newcommand\fverb{\setbox\fverbbox=\hbox\bgroup\verb}
\newcommand\fverbdo{\egroup\medskip\noindent%
			\fbox{\unhbox\fverbbox}\ }
\newcommand\fverbit{\egroup\item[\fbox{\unhbox\fverbbox}]}
\newbox\fverbbox

\makeatletter

\newcommand{\Rmnum}[1]{\expandafter\@slowromancap\romannumeral #1@}
\makeatother

\begin{document}
\title{\mbox{}\\[10pt]
$Z$ boson radiative decays to a $P$-wave quarkonium at NNLO and LL accuracy}

\author{Wen-Long Sang~\footnote{wlsang@swu.edu.cn}}
 \affiliation{School of Physical Science and Technology, Southwest University, Chongqing 400700, China\vspace{0.2cm}}

\author{De-Shan Yang~\footnote{yangds@ucas.ac.cn}}
 \affiliation{School of Physical Sciences, University of Chinese Academy of Sciences, Beijing 100049, China\vspace{0.2cm}}

\author{Yu-Dong Zhang~\footnote{zhangyudong@email.swu.edu.cn}}
 \affiliation{School of Physical Science and Technology, Southwest University, Chongqing 400700, China\vspace{0.2cm}}

\date{\today}
\begin{abstract}

In this work, we study the radiative decay of $Z$ boson to a $P$-wave quarkonium $H$ in association with a photon, where
$H$ can be $\chi_{QJ}$, $h_Q$ with $Q=c,b$ and $J=0,1,2$. The helicity amplitudes and the unpolarized decay widths are evaluated up to QCD next-to-next-to-leading order (NNLO)
within the framework of  nonrelativistic QCD (NRQCD).
For the first time,  we check the NRQCD factorization for $h_Q$ exclusive production at two-loop order.
The leading logarithms (LL) of $m_Z^2/m_Q^2$ in the leading-twist short-distance coefficients, which may potentially ruin the perturbative convergence,
are resummed to all orders of $\alpha_s$ by employing the light-cone factorization.
We find the radiative corrections are considerable  for $\chi_{Q2}$ and $h_Q$ productions, while are moderate or even minor for other channels.
We also notice that the LL resummation can change the leading-order predictions for decay widths by more than 25\% for $\chi_{c0,2}$ and $h_c$ productions, and
by around 50\% for $\chi_{c1}$ production.
However, effects of the LL resummation on the next-to-leading-order and NNLO predictions are notably mitigated.
Some phenomenological explorations are also performed.

\end{abstract}

\maketitle
\section{introduction}

The radiative decay of $Z$ boson to a quarkonium serves to an ideal platform to study
the interplay of the perturbative and nonperturbative nature of QCD.
To data,  experimentalists have made many endeavors to search for such processes~\cite{ATLAS:2015vss,ATLAS:2015xkp,CMS:2018fzh}, yet
 failed to find any signals.
In recent years, several high-luminosity lepton colliders are proposed, such as
{\tt ILC}~\cite{Baer:2013cma}, {\tt FCC-ee}~\cite{FCC:2018byv} and {\tt CEPC}~\cite{CEPCStudyGroup:2018ghi},
which are planed to run at $Z$ mass pole for a period of time. Undoubtedly, tremendous $Z$ bosons
will be accumulated.
Thus it will provide more opportunities to probe these rare decay processes.

The exclusive processes $Z\to {\rm quarkonium}+\gamma$ have been extensively studied on the theoretical side.
The computation on these processes can date back to the earlier 1980s by the authors in Ref.~\cite{Guberina:1980dc}.
In Ref.~\cite{Luchinsky:2017jab}, these processes have also been studied at lowest order in $\alpha_s$ and $v^2$ in both the nonrelativisitic QCD (NRQCD)~\cite{Bodwin:1994jh} and light-cone (LC) factorization formalisms~\cite{Lepage:1980fj,Chernyak:1983ej}, where $v$ represents the typical velocity of the heavy quark in the quarkonium rest frame.
In Ref.~\cite{Wang:2013ywc}, the analytic expressions of the amplitudes for $Z\to {\rm quarkonium}+\gamma$ were obtained in the leading-power LC approximation at NLO in $\alpha_s$.
In Ref.~\cite{Huang:2014cxa}, calculations of the rates for $Z\to V+\gamma$, where $V$ signifies a vector quarkonium $J/\psi$ or $\Upsilon$, were presented.
The calculations were accurate up to the leading-power LC approximation at NLO in $\alpha_s$ and $v$.
Shortly afterwards, the decay rates for $Z\to V+\gamma$ were restudied in Ref.~\cite{Grossman:2015cak}, where
the resummation of the leading logarithms (LL) of $m_Z^2/m_Q^2$, with
$m_Z$ and $m_Q$ being the masses of the $Z$ boson and heavy quark $Q$ respectively, were carried out.
In Ref.~\cite{Bodwin:2017pzj}.  the authors have further considered the resummation of logarithms of $m_Z^2/m_Q^2$ for the $\mathcal{O}(\alpha_s)$
corrections as well as   the $\mathcal{O}(v^2)$ corrections.
Very recently,  the decay rates for $Z\to \Upsilon(nS)+\gamma$ have been calculated up to NLO in $\alpha_s$ based on the NRQCD, which are proposed
to determine the $Zb\bar{b}$ coupling~\cite{Dong:2022ayy}.
As relevant studies, the cross sections of $e^+e^-\to{\rm charmonium}+\gamma$ at $Z$ factories have been computed
at LO and NLO in $\alpha_s$ in Ref.~\cite{Chen:2013mjb}  and in Ref.~\cite{Chen:2013itc} respectively, and the cross sections of
$e^+e^-\to{\rm bottomonium}+\gamma$ at $Z$ factories have been studied in Ref.~\cite{Sun:2014kva}.

In this work, we study the processes of $Z$ boson radiative decays to a $P$-wave quarkonium, i.e., $Z\to H+\gamma$, where
$H$ can be $\chi_{QJ}$, $h_Q$ with $Q=c,b$ and $J=0,1,2$.
Based on the NRQCD factorization and helicity formulas, we compute the various helicity amplitudes at next-to-next-to-leading order (NNLO) in $\alpha_s$ and leading order (LO) in $v$.
Since the two typical energy scales $m_Q$ and $m_Z$ involved in these processes are widely separated, the NRQCD short-distance coefficients (SDCs)  receive contribution from large logarithms of $m_Z^2/m_Q^2$,
which may potentially ruin the perturbative convergence in $\alpha_s$.
Fortunately, it was pointed out~\cite{Jia:2008ep} that the NRQCD SDCs can be refactorized in the framework of LC formalism, in which
the large logarithms can be resummed by employing the celebrated Efremov-Radyushkin-Brodsky-Lepage (ERBL) equation~\cite{Lepage:1980fj,Efremov:1979qk}.
We will carry out the LL resummation for the leading-twist helicity amplitudes.
Thus our computation for $Z\to H+\gamma$ will be at NNLO in $\alpha_s$ at fixed-order accuracy, meanwhile at all orders in $\alpha_s$ at LL accuracy.

The rest of the paper is organized as follows.
In Sec.~\ref{sec-gen-for}, we employ the helicity amplitude
formalism to analyze the $Z\to H+\gamma$ processes, and build the
polarized and unpolarized decay rates out of various helicity amplitudes.
In Sec.~\ref{sec-NRQCD},
we factorize the helicity amplitudes by employing the NRQCD factorization formalism,
and parameterize the corresponding SDCs through NNLO in $\alpha_s$.
The key technical ingredients of extracting the SDCs affiliated with each helicity amplitude through $\alpha_s^2$
are sketched, and values of the SDCs at various perturbative levels are presented.
The corresponding details about constructions of all the helicity projectors are presented in Appendix~\ref{appendix-helicity-projectors}.
We devote Sec.~\ref{sec-LL}  to the LC factorization for the leading-twist helicity SDCs.  The resummation of the LL is formulated and explicitly carried out.
In Sec.~\ref{sec-phen}, a detailed phenomenological analysis is performed. Finally, we summarize in Sec.~\ref{sec-summary}.

\section{the general formula~\label{sec-gen-for}}

It is convenient to employ the helicity amplitude formalism to analyze the hard
exclusive production process.
The differential decay width of $Z$ boson with polarization (along the $z$-axis) $S_z$ into a quarkonium $H$ and a photon,
the helicities of which are $\lambda_1$ and $\lambda_2$, respectively, can be expressed as~\cite{Haber:1994pe,Jacob:1959at}
\bqa\label{eq-gen-rate-helicity}
\frac{\mathrm{d}\Gamma}{\mathrm{d}\cos\theta}(Z(S_{z}) \rightarrow H(\lambda_1)+\gamma(\lambda_2)) =\frac{|\mathbf{P}|}{16 \pi m_{Z}^{2}}\left|d_{S_{z}, \lambda_1-\lambda_2}^{1}(\theta)\right|^{2}\left|A_{\lambda_1, \lambda_2}^{H}\right|^{2},
\eqa
where $\mathbf{P}$ denotes the spatial components of the $H$ momentum,
$\mathcal{A}_{\lambda_1, \lambda_2}^{H}$ represents the amplitude corresponding to the helicity configuration ($\lambda_1$, $\lambda_2$),
and $d^1_{S_z,\lambda_1-\lambda_2}(\theta)$ is the Wigner function.
Here, $\theta$ is the angle between the direction of $\mathbf{P}$ and the $z$-axis.
Note that the constraint, $\lambda_{1}-\lambda_{2}\leq 1$, is guaranteed by the angular momentum conservation.
The magnitude of the spatial momentum $|{\bf P}|$ is readily determined via
\beq
|{\bf P}|=\frac{\lambda^{1/2}(m_Z^2,m_H^2,0)}{2m_Z}=\frac{m_Z^2-m_H^2}{2m_Z},
\eeq
where $m_H$ denotes the mass of the quarkonium $H$, and the K\"allen function is defined via $\lambda(x,y,z)=x^2+y^2+z^2-2xy-2xz-2yz$.

Integrating over the polar angle $\theta$ and averaging over the polarization of $Z$,
we finally obtain the integrated decay width of $Z\to H+\gamma$  for the helicity configuration ($\lambda_{1}$,$\lambda_{2}$) as
\begin{equation}\label{eq-gen-rate-helicity-int}
\Gamma(Z \rightarrow H(\lambda_1)+\gamma(\lambda_{2}))=\frac{|\mathbf{P}|}{24\pi m_{Z}^{2}}\left|A_{\lambda_1,\lambda_2}^{H}\right|^{2}.
\end{equation}
Thanks to the parity invariance~\cite{Haber:1994pe}, we have the following relations,
\bqa\label{eq-helicity-parity-invariance}
A^{\chi_{QJ}}_{\lambda_1,\lambda_2}=(-1)^J A^{\chi_{QJ}}_{-\lambda_1,-\lambda_2},
\quad
A^{h_Q}_{\lambda_1,\lambda_2}=A^{h_Q}_{-\lambda_1,-\lambda_2}.
\eqa
Thus the number of independent helicity amplitudes for $\chi_{Q0}$, $\chi_{Q1}$, $\chi_{Q2}$
and $h_Q$ production can be reduced to one, two, three, and two, respectively.

In the limit of $m_Z\gg m_Q$, the helicity amplitude $A_{\lambda_1,\lambda_2}^H$ satisfies the asymptotic behavior
\bqa\label{eq-helicity-selection-rule}
A_{\lambda_1,\lambda_2}^H \propto r^{1+|\lambda_1|},
\eqa
where $r$ is defined via $r=m_Q/m_Z$. One power of $r$ in Eq.~(\ref{eq-helicity-selection-rule}) originates from the large
momentum transfer which are required for the heavy-quark pair to form the heavy quarkonium with small relative
momentum and the other powers arise from the helicity selection rule in perturbative QCD~\cite{Chernyak:1980dj,Brodsky:1981kj}.

In terms of the independent helicity amplitudes, the unpolarized decay widths can be explicitly written as
\begin{subequations}
\begin{eqnarray}\label{eq-gen-rate-helicity-explicit}
\Gamma(Z\to \chi_{Q0}+\gamma)&=&
\frac{1}{3}\frac{1}{2m_Z}\frac{1}{8\pi}\frac{2|{\bf P}|}{m_Z}\bigg(2| A_{0,1}^{\chi_{Q0}}|^2\bigg),\\
\Gamma(Z\to \chi_{Q1}+\gamma)&=&
\frac{1}{3}\frac{1}{2m_Z}\frac{1}{8\pi}\frac{2|{\bf P}|}{m_Z}\bigg(2| A_{1,1}^{\chi_{Q1}}|^2
+2| A_{0,1}^{\chi_{Q1}}|^2\bigg),\\
\Gamma(Z\to \chi_{Q2}+\gamma)&=&
\frac{1}{3}\frac{1}{2m_Z}\frac{1}{8\pi}\frac{2|{\bf P}|}{m_Z}\bigg(2| A_{2,1}^{\chi_{Q2}}|^2
+2|A_{1,1}^{\chi_{Q2}}|^2
+2|A_{0,1}^{\chi_{Q2}}|^2\bigg),\\
\Gamma(Z\to h_{Q}+\gamma)&=&
\frac{1}{3}\frac{1}{2m_Z}\frac{1}{8\pi}\frac{2|{\bf P}|}{m_Z}\bigg(2| A_{1,1}^{h_Q}|^2
+2| A_{0,1}^{h_Q}|^2\bigg).
\end{eqnarray}
\end{subequations}

The main task of this work remains to compute the helicity amplitudes.
$Z$ boson interacts with quark-anti-quark pair through the tree-level weak interaction as
\bqa\label{eq-z-qq}
i\mathcal{L}_{ZQ\bar{Q}}=i\frac{g}{4 c_W}\bar{Q}\gamma^\mu(g_V-g_A\gamma_5)QZ_{\mu},
\eqa
where $g$ is the weak coupling in $SU(2)_L\times U(1)_Y$ electro-weak gauge theory, $g_V=1-8s^2_W/3$ and $g_A=1$ for the up-type quark,
and $g_V=-1+4s^2_W/3$ and
$g_A=-1$ for the down-type quark. Here we have defined $s_W\equiv \sin\theta_W$, and $c_W\equiv \cos\theta_W$,
where $\theta_W$ signfies the Weinberg angle.

$Z$ boson can decay to $\chi_{QJ}+\gamma$ only through the vectorial interaction, and
decay to $h_Q+\gamma$ only through the axial-vectorial interaction. Therefore, for simplicity, it is convenient to explicitly extract the electro-weak coupling from the helicity amplitudes as
\begin{subequations}
\begin{eqnarray}\label{eq-helicity-amplitude-1}
 A_{\lambda_1,\lambda_2}^{\chi_{QJ}}&=&\frac{gg_Vee_Q}{4c_W}\mathcal{A}_{\lambda_1,\lambda_2}^{\chi_{QJ}},\\
 A_{\lambda_1,\lambda_2}^{h_Q}&=&\frac{gg_Aee_Q}{4c_W}\mathcal{A}_{\lambda_1,\lambda_2}^{h_{Q}}.
\end{eqnarray}
\end{subequations}


\section{Framework of NRQCD computation\label{sec-NRQCD}}
\subsection{NRQCD factorization}

According to the NRQCD formalism~\cite{Bodwin:1994jh},
the helicity amplitude ${\mathcal A}_{\lambda_1,\lambda_2}^{H}$ can be
factorized into
\beq\label{eq-nrqcd}
{\mathcal A}_{\lambda_1,\lambda_2}^{H}=\sqrt{2m_H} \mathcal{C}_{\lambda_1,\lambda_2}^H
\frac{\langle\mathcal{O}\rangle_H}{\sqrt{2N_c}2m_Q^2},
\eeq
where $\mathcal{C}_{\lambda_1,\lambda_2}^H$ signify the dimensionless SDCs, $N_c=3$ is the number of the color,  and the NRQCD long-distance matrix elements (LDMEs) are defined via
\begin{subequations}
\begin{eqnarray}\label{eq-NRQCD-operators}
\langle \mathcal{O} \rangle_{\chi_{QJ}}&\equiv&\langle \chi_{QJ}|
\psi^{\dagger}{\cal K}_{^3P_J} \chi |0\rangle,\\
\langle \mathcal{O}\rangle_{h_Q}&\equiv&\langle h_{Q}|
\psi^{\dagger}{\cal K}_{^1P_1} \chi |0\rangle,
\end{eqnarray}
\end{subequations}
where $\psi^\dagger$ and $\chi$ denote the Pauli spinor fields creating a heavy quark and antiquark in NRQCD, respectively, and
\begin{subequations}
\begin{eqnarray}
{\cal K}_{^3P_{0}}&=&\frac{1}{\sqrt{3}}\left(-\frac{i}{2}\overleftrightarrow{{\bf
D}}\cdot {\bm\sigma}\right),\\
{\cal K}_{^3P_1}&=&\frac{1}{\sqrt{2}}\left(-\frac{i}{2}\overleftrightarrow{{\bf
D}}\times\bm{\sigma}\right)\cdot \bm{\epsilon}_{\chi_{Q1}},\\
{\cal K}_{^3P_2}&=&-{i\over 2}\overleftrightarrow{D}^{(i}\sigma^{j)}\epsilon^{ij}_{\chi_{Q2}},\\
{\cal K}_{^1P_{1}}&=& -\frac{i}{2}\overleftrightarrow{{\bf D}}\cdot \bm{\epsilon}_{h_{Q}},
\end{eqnarray}
\end{subequations}
with $ \epsilon_{\chi_{Q1}}$,  $\epsilon_{h_{Q}}$, and $\epsilon_{\chi_{Q2}}$ representing the polarization vector/tensor of $\chi_{Q1}$,
$h_Q$, and $\chi_{Q2}$ respectively.  Exploiting the heavy quark spin symmetry, we can make the following approximations
\begin{eqnarray}\label{eq-NRQCD-operators}
\langle\mathcal{O}\rangle_{\chi_{Q0}}\approx
\langle\mathcal{O}\rangle_{\chi_{Q1}}\approx \langle\mathcal{O}\rangle_{\chi_{Q2}}
\approx \langle\mathcal{O}\rangle_{h_Q}.
\end{eqnarray}

In Eq.~(\ref{eq-nrqcd}), the factor $\sqrt{2m_H}$ appears on the right side because
we adopt relativistic normalization for the quarkonium $H$, but we use conventional
nonrelativistic normalization for the LDMEs.
In this work, we will not compute the relativistic corrections,
therefore it is reasonable to take the approximation $m_H\approx 2m_Q$.

Through Eq.~(\ref{eq-helicity-selection-rule}) and Eq.~(\ref{eq-nrqcd}),
we can readily deduce the helicity selection rule for the SDCs
\bqa\label{eq-helicity-scaling-rule-1}
{\mathcal C}_{\lambda_1,\lambda_2}^H\propto r^{|\lambda_1|}
\eqa
by noting that $\langle\mathcal{O}\rangle_H\propto m_Q^{5/2}$.

The SDCs are insensitive to the nonperturbative hadronization effects,
therefore they can be determined with the aid of the standard perturbative matching technique. That is, by
replacing the physical $H$ meson with a fictitious onium composed of  a free
$Q\bar{Q}$ pair, carrying
the quantum number $^3P_J$ for $\chi_{QJ}$ and $^1P_1$ for $h_Q$,
we compute both sides of Eq.~(\ref{eq-nrqcd}) .
After that, we are able to solve for the desired SDCs.
For more details,
we refer the readers to Ref.~\cite{Sang:2020fql}.

It is convenient to parametrize the SDCs in powers of $\alpha_s$
\bqa\label{eq-sdcs-expand}
\mathcal{C}_{\lambda_1,\lambda_2}^H&=&\mathcal{C}_{\lambda_1,\lambda_2}^{H,(0)}\bigg[1+\frac{\alpha_s}{\pi}
\mathcal{C}_{\lambda_1,\lambda_2}^{H,(1)}
+\frac{\alpha_s^2}{\pi^2}\bigg(\frac{\beta_0}{4}\ln\frac{\mu_R^2}{m_Q^2}\mathcal{C}_{\lambda_1,\lambda_2}^{H,(1)}
+\gamma_H \ln\frac{\mu_\Lambda}{m_Q}\nn\\
&&~~~~~~~~~~~+ \mathcal{C}_{{\rm reg},\lambda_1,\lambda_2}^{H,(2)}+ \mathcal{C}_{{\rm nonreg},\lambda_1,\lambda_2}^{H,(2)}\bigg)\bigg]
+\mathcal{O}(\alpha_s^3),
\eqa
where $\mu_R$ and $\mu_\Lambda$ signify the renormalization scale and factorization scale, respectively,
$\beta_0=(11/3)C_A-(4/3)T_Fn_f$ is the one-loop coefficient of the QCD $\beta$ function,
where $n_f$ is the number of active quark flavors.
The explicit $\ln\mu_R^2$ term is deduced from the renormalization-group invariance.
$\gamma_H$ represent the anomalous dimensions associated with the NRQCD
bilinear currents carrying the quantum numbers $^3P_J$ or $^1P_1$,  the expressions of which
read~\cite{Hoang:2006ty}
\begin{subequations}\label{eq-ano-dim}
\begin{eqnarray}
\gamma_{^3P_0}&=&-\pi^2\bigg(\frac{C_AC_F}{6}+\frac{2C_F^2}{3}\bigg),\\
\gamma_{^3P_1}&=&-\pi^2\bigg(\frac{C_AC_F}{6}+\frac{5C_F^2}{12}\bigg),\\
\gamma_{^3P_2}&=&-\pi^2\bigg(\frac{C_AC_F}{6}+\frac{13C_F^2}{60}\bigg),\\
\gamma_{^1P_1}&=&-\pi^2\bigg(\frac{C_AC_F}{6}+\frac{C_F^2}{3}\bigg).
\end{eqnarray}
\end{subequations}
The occurrence of  $\ln\mu_\Lambda$ is required by the NRQCD factorization. According to the factorization, the $\mu_\Lambda$ dependence in the SDCs should
be thoroughly canceled by that in the LDMEs.
As illustrated in Fig.~\ref{fig-feynman-diagram}, we classify the Feynman diagrams into  `regular' part and `nonregular' part  at $\mathcal{O}(\alpha_s^2)$.
Correspondingly,  $\mathcal{C}_{{\rm reg},\lambda_1,\lambda_2}^{H,(2)}$
and $\mathcal{C}_{{\rm nonreg},\lambda_1,\lambda_2}^{H,(2)}$ in Eq.~(\ref{eq-sdcs-expand})  represent contributions from the `regular' Feynman diagrams
and the `nonregular' Feynman diagrams respectively.

\subsection{SDCs through $\mathcal{O}(\alpha_s^2)$}

\begin{figure}[htbp]
	\centering
	\includegraphics[width=1\textwidth]{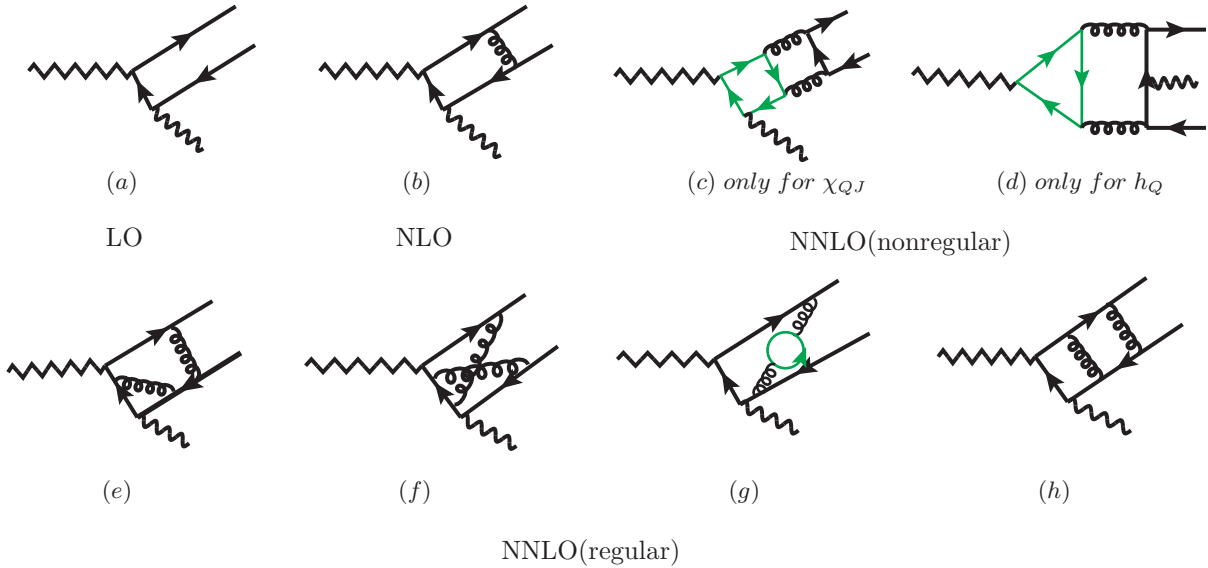}
	\caption{Some representative Feynman diagrams for the process $Z\to\chi_{QJ}(h_{Q})+\gamma$ up to $\mathcal{O}(\alpha_s^2)$.
	\label{fig-feynman-diagram}}
\end{figure}
The quark-level Feynman diagrams and Feynman amplitudes are generated using {\tt FeynArts}~\cite{Hahn:2000kx}.
Employing the color and spin projectors followed by enforcing spin-orbit coupling,
we obtain the hadron-level amplitudes order by order in $\alpha_s$ with the aid of the packages {\tt FeynCalc}~\cite{Shtabovenko:2016sxi} and {\tt FormLink}~\cite{Feng:2012tk}.
To evaluate the helicity amplitudes, we employ the technique of helicity projection.
The concrete expressions of all the helicity projectors are presented in Appendix.~\ref{appendix-helicity-projectors}.

It is well known that, in dimensional regularization, the anticommutation relation $\{\gamma^\mu,\gamma_5\}$
and the cyclicity of Dirac trace can not be satisfied simultaneously. In practical computation,
the naive-$\gamma_5$ scheme~\cite{Korner:1991sx}, which keeps the anticommutation relation $\{\gamma^\mu,\gamma_5\}$, is frequently applied. In this scheme,
spurious anomaly, which spoils chiral symmetry and hence gauge invariance can be avoided.
Due to the lack of the cyclicity of the trace, one must fix a reading point for a fermion loop with odd number of $\gamma_5$.
In our work, we will select the vertex of $Z$ boson as the reading point.

Then, it is straightforward to obtain the LO helicity SDCs:
\begin{subequations}
\begin{eqnarray}
\mathcal{C}_{0,1}^{\chi_{Q0},(0)}&=&\frac{2\sqrt{2}(1-12r^2)}{1-4r^2},\\
\mathcal{C}_{1,1}^{\chi_{Q1},(0)}&=&-\frac{8\sqrt{3}r}{1-4r^2},\quad \quad
\mathcal{C}_{0,1}^{\chi_{Q1},(0)}=\frac{4\sqrt{3}}{1-4r^2},\\
\mathcal{C}_{2,1}^{\chi_{Q2},(0)}&=&-\frac{16\sqrt{6}r^2}{1-4r^2},\quad \quad
\mathcal{C}_{1,1}^{\chi_{Q2},(0)}=\frac{8\sqrt{3}r}{1-4r^2},\quad \quad
\mathcal{C}_{0,1}^{\chi_{Q2},(0)}=-\frac{4}{1-4r^2},\\
\mathcal{C}_{1,1}^{h_Q,(0)}&=&4\sqrt{6}r,  \quad \quad
\mathcal{C}_{0,1}^{h_Q,(0)}=-2\sqrt{6}.
\end{eqnarray}
\end{subequations}

Once beyond the LO, we adopt the standard shortcut to directly extract the SDCs,
i.e., compute the hard region in the context of strategy of region~\cite{Beneke:1997zp}.
Utilizing the packages {\tt Apart}~\cite{Feng:2012iq} and {\tt FIRE}~\cite{Smirnov:2014hma},
we can further reduce the loop integrals into linear combinations of master integrals (MIs).
Finally, we end up with 6 one-loop MIs, which are computed using {\tt Package-X}~\cite{Patel:2015tea},
and roughly 320 two-loop MIs, the evalutaion of which is a challenging work.
Fortunately, a powerful new algorithm, dubbed Auxiliary Mass Flow (AMF),
has recently been pioneered by Liu and Ma~\cite{Liu:2017jxz,Liu:2020kpc,Liu:2021wks,Liu:2022mfb}.
Its main idea is to set up differential equations with respect to an auxiliary mass variable, with the vacuum bubble diagrams as the boundary conditions.
Remarkably, these differential equations can be solved iteratively with very high numerical precision.
In this work, we utilize the newly released package {\tt AMFlow}~\cite{Liu:2022chg} to compute all the two-loop MIs.
After implementing the on-shell renormalization scheme for the heavy quark mass and field strength~\cite{Fael:2020bgs}, and
the $\overline{\mathrm{MS}}$ renormalization scheme for the QCD coupling, the
UV poles are exactly canceled, while an uncancelled
single IR pole still remains. This symptom is a common feature specific to the NRQCD factorization,
which have been encountered many times in NNLO perturbative calculations involving
quarkonium.  This IR pole can be factored into the NRQCD LDME, so that the NRQCD
SDCs become IR finite.
We have numerically verified that the coefficient of the remaining IR pole equal
to one-quarter of the anomalous dimension in Eq.~(\ref{eq-ano-dim})  with high precision, as
required by the NRQCD factorization.

The analytic expressions of $\mathcal{C}_{\lambda_1,\lambda_2}^{H,(1)}$ can be readily obtained.
Instead of presenting the cumbersome expressions, here we merely present their asymptotic expansions in $r\to 0$:
\begin{subequations}\label{eq-SDCs-1}
\begin{eqnarray}
\mathcal{C}_{0,1}^{\chi_{Q0},(1)}&=&\frac{1}{3}(2\ln2-1)\ln(-r^2+i\epsilon)+\frac{\ln^22}{3}+3\ln 2-\frac{\pi^2}{9},\\
\mathcal{C}_{1,1}^{\chi_{Q1},(1)}&=&\frac{1}{3}(2\ln2-3)\ln(-r^2+i\epsilon)+\frac{\ln^22}{3}-\frac{\ln 2}{3}-\frac{\pi^2}{9}-2\\
\mathcal{C}_{0,1}^{\chi_{Q1},(1)}&=&\frac{1}{3}(2\ln2-3)\ln(-r^2+i\epsilon)+\frac{\ln^22}{3}-\frac{\ln 2}{3}-\frac{\pi^2}{9}-\frac{7}{3},\\
\mathcal{C}_{2,1}^{\chi_{Q2},(1)}&=&\frac{4}{3}(2\ln2-1)\ln(-r^2+i\epsilon)+\frac{4\ln^2 2}{3}+\frac{16\ln2}{3}-\frac{4\pi^2}{9}-\frac{8}{3},\\
\mathcal{C}_{1,1}^{\chi_{Q2},(1)}&=&\frac{1}{3}(2\ln2+1)\ln(-r^2+i\epsilon)+\frac{\ln^2 2}{3}-\frac{5\ln2}{3}-\frac{\pi^2}{9},\\
\mathcal{C}_{0,1}^{\chi_{Q2},(1)}&=&\frac{1}{3}(2\ln2-1)\ln(-r^2+i\epsilon)+\frac{\ln^2 2}{3}-\ln 2-\frac{\pi^2}{9}-2,\\
\mathcal{C}_{1,1}^{h_Q,(1)}&=&\frac{4}{3}\ln2\ln(-r^2+i\epsilon)+\frac{2\ln^2 2}{3}+\frac{8\ln 2}{3}-\frac{2\pi^2}{9}-1,  \\
\mathcal{C}_{0,1}^{h_Q,(1)}&=&\frac{1}{3}(2\ln2-1)\ln(-r^2+i\epsilon)+\frac{\ln^2 2}{3}-\ln 2-\frac{\pi^2}{9}-\frac{4}{3},
\end{eqnarray}
\end{subequations}
where the real parts of the results in Eqs.~(\ref{eq-SDCs-1}a-f) are consistent with these in Ref.~\cite{Sang:2009jc}~\footnote{In Ref.~~\cite{Sang:2009jc},
only the cross sections of $e^+e^-\to \chi_{cJ}+\gamma$ are presented, where the imaginary parts of the amplitudes are ignored.},
and the results in Eqs.~(\ref{eq-SDCs-1}a,c,f,h) are consistent
with these in Ref.~\cite{Wang:2013ywc}.

It is rather challenging for us to  analytically compute $\mathcal{C}_{\lambda_1,\lambda_2}^{H,(2)}$, so we turn to numerically evaluate their values.
To perform the numerical computation, we take $m_Z=91.1876$ GeV from the latest particle data group (PDG)~\cite{ParticleDataGroup:2020ssz},
and the pole masses of the charm quark and bottom quark to be $m_c = 1.69$ GeV and $m_b=4.80$ GeV, which are converted from the $\overline{\rm MS}$ masses $\bar{m}_c(\bar{m}_c)=1.28$ GeV and $\bar{m}_b(\bar{m}_b)=4.18$ GeV~\cite{ParticleDataGroup:2020ssz} at two-loop level by use of the package {\tt RunDec}~\cite{Chetyrkin:2000yt}.

We tabulate  the results of the SDCs $\mathcal{C}^{H,(1)}_{\lambda_1,\lambda_2}$,
$\mathcal{C}_{{\rm reg},\lambda_1,\lambda_2}^{H,(2)}$ and $\mathcal{C}_{{\rm nonreg},\lambda_1,\lambda_2}^{H,(2)}$ in Tab.~\ref{tab-sdcs-mc-1.69} for charmonium production, and  in Tab.~\ref{tab-sdcs-mb-4.80} for bottomonium production~\footnote{Since weak interaction in Standard Model (SM) is a chiral gauge theory, the gauge anomaly should be avoided for physical processes. To satisfy the condition of anomaly free, when evaluating $\mathcal{C}_{{\rm nonreg},\lambda_1,\lambda_2}^{h_Q,(2)}$ which corresponds to contribution
from Fig.~\ref{fig-feynman-diagram}d,  we have included all six flavor quark loops.}.
For the sake of reference, we explicitly remain the $n_l$, $n_c$ and $n_b$ dependence in the SDCs, where $n_l$ denotes the number of the light quarks,
and $n_c=1$ and $n_b=1$ signify the numbers of the charm quark and bottom quark respectively.

\begin{table}[!htbp]\small
	\caption{NRQCD predictions to the various helicity SDCs defined in  Eq.(\ref{eq-sdcs-expand}) for charmonium production.
	For simplicity, we define the symbols $f_{1}\equiv\frac{g_V^d}{g_V^u}=-\frac{3-4 S_W^2}{3-8 S_W^2}$ and $f_{2}\equiv\frac{g_V^u-g_V^d}{g_V^u}=\frac{6-12 S_W^2}{3-8 S_W^2}$, where $g_V^u$ and $g_V^d$ correspond to the values of $g_V$ for
	up-type quark and down-type quark respectively.}
	\label{tab-sdcs-mc-1.69}
	\setlength{\tabcolsep}{0.4mm}
	\centering
\resizebox{\textwidth}{!}{
	\begin{tabular}{|c|c|c|c|c|}
		\hline
		{$H$}
		& $(\lambda_1,\lambda_2)$
		& $\mathcal{C}_{\lambda_1,\lambda_2}^{(1)}$
		&$\mathcal{C}_{{\rm reg},\lambda_1,\lambda_2}^{(2)}$
		&$\mathcal{C}_{{\rm nonreg},\lambda_1,\lambda_2}^{(2)}$\\
		\hline
		\multirow{2}*{$\chi_{c0}$}
		& \multirow{2}*{(0,1)}
		& \multirow{2}*{$0.09+0.43i$}
		& $-2.00-2.21 i-(2.34-1.29 i)n_l$
		& $(6.43-7.29 i)n_c-(1.57-3.71 i)f_{1}n_b $\\
		&
		&
		&$-(1.12-1.28 i)n_c-(0.58-1.25 i)n_b$
		& $+(9.49-6.37 i)f_{2}$\\
		\hline
		\multirow{4}*{$\chi_{c1}$}
		&  \multirow{2}*{(1,1)}
		& \multirow{2}*{$1.11-1.69i$}
		& $-29.65+22.87 i-(0.32+0.92 i)n_l$
		& $-(3.01-1.60i)n_c+(0.72-0.78 i)f_{1}n_b$\\
		&
		&
		&$-(0.09+0.92i)n_c-(0.03+0.93 i)n_b$
		& $-(3.43+0.04 i)f_{2}$  \\
		\cline{2-5}
		&  \multirow{2}*{(0,1)}
		& \multirow{2}*{$0.77-1.68i$}
		& $-28.51+24.13 i-(0.53+0.74 i)n_l$
		& $-(3.67-2.73 i)n_c+(0.77-1.30 i)f_{1}n_b$\\
		&
		&
		& $-(0.29+0.74 i)n_c-(0.23+0.76 i)n_b$
		&$-(4.66-0.89i)f_{2}$ \\
		\hline
		\multirow{6}*{$\chi_{c2}$}
		&  \multirow{2}*{(2,1)}
		& \multirow{2}*{$-6.83+1.62i$}
		& $240.12-37.67 i+(0.06+1.67 i)n_l$
		& $(181.76-21.31 i)n_c-(88.77-40.30 i)f_{1} n_b$\\
		&
		&
		&$-(1.45-1.67 i)n_c-(2.93-1.68 i)n_b$
		&$+(173.42-2.68 i)f_{2}$  \\
		\cline{2-5}
		&  \multirow{2}*{(1,1)}
		& \multirow{2}*{$-8.44+2.50i$}
		& $10.19-53.70 i-(0.22-2.43 i)n_l$
		& $(3.35-0.65 i)n_c-(1.59-0.79 i)f_{1}n_b$\\
		&
		&
		& $-(1.84-2.43 i)n_c-(3.52-2.43 i)n_b$
		&$+(3.37-0.29i)f_{2}$\\
		\cline{2-5}
		&  \multirow{2}*{(0,1)}
		& \multirow{2}*{$-4.69+0.43i$}
		& $-6.96-8.92 i-(0.84-1.30 i)n_l$
		& $(5.23-2.15 i)n_c-(1.89-1.85 i)f_{1}n_b$\\
		&
		&
		& $-(1.60-1.29 i)n_c-(2.40-1.25 i)n_b$
		&$+(5.10-1.31)f_{2}$ \\
		\hline
		\multirow{4}*{$h_c$}
		&  \multirow{2}*{(1,1)}
		& \multirow{2}*{$-8.39+2.89i$}
		& $10.52-64.86 i-(0.54-2.62 i)n_l$
		& \multirow{2}*{$1.66+1.62 i$}\\
		&
		&
		& $-(1.85-2.63 i)n_c-(3.46-2.64 i)n_b$
		&  \\
		\cline{2-5}
		& \multirow{2}*{(0,1)}
		& \multirow{2}*{$-4.01+0.43i$}
		& $-(9.98+7.77 i)-(1.20-1.29 i)n_l$
		& \multirow{2}*{$1.75+1.56 i$}\\
		&
		&
		& $-(1.57-1.29 i)n_c-(2.15-1.25 i)n_b$
		& \\
		\hline
	\end{tabular}
}
\end{table}
\begin{table}[!htbp]\small
\caption{NRQCD predictions to the various helicity SDCs for bottomonium production.
	For simplicity, we define the symbols
$\bar{f}_{1}\equiv\frac{g_V^u}{g_V^d}=-\frac{3-8 S_W^2}{3-4 S_W^2}$,
$\bar{f}_{2}\equiv\frac{2g_V^d-2g_V^u}{g_V^d}=\frac{12-24 S_W^2}{3-4 S_W^2}$.}
\label{tab-sdcs-mb-4.80}
\setlength{\tabcolsep}{0.4mm}
\centering
\resizebox{\textwidth}{!}{
\begin{tabular}{|c|c|c|c|c|}
	\hline
	{$H$}
	&($\lambda_1,\lambda_2$)
	&$\mathcal{C}_{\lambda_1,\lambda_2}^{(1)}$
	&$\mathcal{C}_{{\rm reg},\lambda_1,\lambda_2}^{(2)}$
	&$\mathcal{C}_{{\rm nonreg},\lambda_1,\lambda_2}^{(2)}$\\
	\hline
	\multirow{2}*{$\chi_{b0}$}
	& \multirow{2}*{(0,1)}
	&\multirow{2}*{$0.35+0.58 i$}
	&$-4.41-6.16 i-(1.47-1.15 i)n_l$
	&$-(8.02-11.05 i)\bar{f}_{1} n_c+(1.83-5.06 i)n_b$\\
	&
	&
	&$-(0.35-1.14i)n_c-(0.22-1.10 i)n_b$
	&$+(5.06-4.40i)\bar{f}_{2}$\\
	\hline
	\multirow{4}*{$\chi_{b1}$}
	& \multirow{2}*{(1,1)}
	&\multirow{2}*{$-0.07-1.65i$}
	&$-18.06+10.67 i-(0.75+0.28 i)n_l$
	&$+(4.83-0.96 i)\bar{f}_{1} n_c-(1.81-1.75i)n_b$\\
	&
	&
	&$-(0.17+0.28i)n_c-(0.50+0.29i)n_b$
	&$-(2.28-0.14i)\bar{f}_{2}$  \\
	\cline{2-5}
	& \multirow{2}*{(0,1)}
	&\multirow{2}*{$-0.42-1.61i$}
	&$-16.52+10.33 i-(0.83+0.10 i)n_l$
	&$+(5.54-2.50 i)\bar{f}_{1}n_c-(1.76-2.47 i)n_b$\\
	&
	&
	&$-(0.25+0.10i)n_c-(0.57+0.13 i)n_b$
	&$-(2.85-0.75i)\bar{f}_{2}$ \\
	\hline
	\multirow{6}*{$\chi_{b2}$}
	& \multirow{2}*{(2,1)}
	&\multirow{2}*{$-5.77+1.63i$}
	&$8.69-27.65 i+(0.98+1.12 i)n_l$
	&$-(47.72-8.15 i)\bar{f}_{1} n_c+(23.15-11.58 i)n_b$\\
	&
	&
	&$+(0.72+1.12 i)n_c-(0.54-1.13 i)n_b$
	& $+(22.90-1.73i)\bar{f}_{2}$  \\
	\cline{2-5}
	&\multirow{2}*{(1,1)}
	&\multirow{2}*{$-6.77+2.50i$}
	&$-20.19-37.37 i+(1.11+1.55 i)n_l$
	&$-(3.88-0.97 i)\bar{f}_{1}n_c+(1.88-0.71 i)n_b$\\
	&
	&
	&$+(0.83+1.55i)n_c-(0.51-1.55 i)n_b$
	&$+(1.89-0.23i)\bar{f}_{2}$\\
	\cline{2-5}
	& \multirow{2}*{(0,1)}
	&\multirow{2}*{$-4.50+0.57i$}
	&$-13.91-13.04 i+(0.02+1.20 i)n_l$
	&$-(4.56-1.86 i)\bar{f}_{1}n_c+(2.20-1.47 i)n_b$\\
	&
	&
	&$+(0.11+1.19i)n_c-(0.73-1.15 i)n_b$
	&$+(2.14-0.70)\bar{f}_{2}$\\
	\hline
	\multirow{4}*{$h_b$}
	& \multirow{2}*{(1,1)}
	&\multirow{2}*{$-6.43+2.85i$}
	&$-(25.69+42.26 i)+(0.87+1.59 i)n_l$
	&\multirow{2}*{$-1.71-1.60 i$}\\
	&
	&
	&$+(0.78+1.59i)n_c-(0.45-1.60 i)n_b$
	& \\
	\cline{2-5}
	&\multirow{2}*{(0,1)}
	&\multirow{2}*{$-3.78+0.54i$}
	&$-(16.35+11.34i)-(0.35-1.17 i)n_l$
	&\multirow{2}*{$-1.76-1.54 i$}\\
	&
	&
	&$-(0.03-1.17i)n_c-(0.69-1.13 i)n_b$
	&   \\
	\hline
\end{tabular}
}
\end{table}

\section{LC factorization for the leading-twist SDCs\label{sec-LL}}

\subsection{The LC factorization~\label{subsec-LL}}
Besides the NRQCD factorization formalism, we can also employ the LC factorization framework
to calculate the decay amplitude for $Z\to H+\gamma$ at the leading twist. By following the spirit of Ref.~\cite{Jia:2008ep},
the LC factorization formula for the SDCs is written as
\begin{eqnarray}
	\mathcal{C}_{0,1}^H(\mu;m_Z,m_Q)=\mathcal{C}^{H,\rm LL_0}_{0,1}\int_0^1 dx T_H(x,\mu) \hat\phi_H(x,\mu)+\mathcal{O}(r^2)\,,
\end{eqnarray}
where $\mathcal{C}^{H,\rm LL_0}_{0,1}$ represents the asymptotic expansion of $\mathcal{C}^{H,(0)}_{0,1}$ in $r\to 0$,
 the hard-kernel $T_H$ and the leading-twist LC distribution amplitude (LCDA) $\hat\phi_H$ are perturbatively calculable around the scale $m_Z$ and $m_Q$, respectively. At LO in $\alpha_s$, we have
\begin{eqnarray}
T_H(x,\mu=m_Z)&=&T_H^{(0)}(x)=\begin{cases}
\frac{1}{4}\left(\frac{1}{1-x}-\frac{1}{x}\right)\,,~~~\text{for~$H=h_Q,\chi_{Q0}$ and $\chi_{Q2}$}\,,\\
\frac{1}{4}\left(\frac{1}{1-x}+\frac{1}{x}\right)\,,~~~\text{for~$H=\chi_{Q1}$\,,}
\end{cases}\\
\hat{\phi}_H(x,\mu=m_Q)&=&\hat{\phi}^{(0)}_H(x)=\begin{cases}
-\frac{1}{2}\delta'(x-1/2)\,,~~~\text{for~$H=h_Q,\chi_{Q0}$ and $\chi_{Q2}$}\,,\\
\delta(x-1/2)\,,~~~~~~~~\text{for~$H=\chi_{Q1}$\,.}
\end{cases}
\end{eqnarray}
We can reproduce the asymptotic expansions of the SDCs $\mathcal{C}_{0,1}^{H,(1)}$ in Eq.~(\ref{eq-SDCs-1}) exactly with the corresponding NLO corrections to $T_H$ and $\hat{\phi}_H$ which have been calculated in \cite{Wang:2013ywc}. More importantly, we can employ the LC factorization to resum the LL terms $(\alpha_s\ln r^2)^n$ in $\mathcal{C}_{0,1}^H$.

\subsection{Resummation of the LL with the ERBL equation~\label{subsec-resum}}
The leading twist LCDAs $\hat\phi_H$ obey the celebrated ERBL equation \cite{Lepage:1979zb, Efremov:1979qk}
\begin{eqnarray}
\mu^2\frac{d}{d\mu^2}\hat\phi_H(x;\mu)=\frac{\alpha_s(\mu)}{2\pi}C_F\int_0^1 V(x,y)\hat\phi_H(y;\mu)\,,
\end{eqnarray}
with the Brodsky-Lepage (BL) kernel
\begin{eqnarray}
V(x,y)=\left[\frac{1-x}{1-y}\left(1+\frac{1}{x-y}\right)\theta(x-y)+\frac{x}{y}\left(1+\frac{1}{y-x}\right)\theta(y-x)\right]_+\,,	
\end{eqnarray}
where the subscript "+" implies the familiar "plus" prescription.

Solving the ERBL equation, we can obtain the SDCs with all the LL terms $(\alpha_s \ln r^2)^n$ resummed. Formally, we have
\begin{eqnarray}\label{eq:LLexp}
\mathcal{C}_{0,1}^{H,\mathrm{LL}}&=&\mathcal{C}_{0,1}^{H,\mathrm{LL}_0} \mathcal{K}_H^{\mathrm{LL}}\,,
\end{eqnarray}
with
\begin{eqnarray}\label{eq:LLsolution}
\mathcal{K}_H^{\rm LL}&=& \int_0^1 dx\, T^{H,(0)}(x)\exp \left(\kappa C_{F} V \star\right) \hat{\phi}^{(0)}_H(x)\,,
\end{eqnarray}
where
\begin{eqnarray}
\kappa \equiv \frac{2}{\beta_{0}} \ln \left(\frac{\alpha_{s}\left(m_Q\right)}{\alpha_{s}\left(m_Z\right)}\right)=-\frac{\alpha_{s}\left(m_Z\right)}{2 \pi} \ln r^2+\beta_{0} \frac{\alpha_{s}^{2}\left(m_Z\right)}{(4 \pi)^{2}} \ln ^{2}r^2+\cdots\,,
\end{eqnarray}
with "$\star$" stands for the convolution
\begin{eqnarray}
\left[V\star \hat\phi\right](x)\equiv \int_0^1 dy V(x,y) \hat\phi(y)\,.
\end{eqnarray}

It is well known that the BL kernel has the eigen-system that
\begin{eqnarray}
\int_0^1 V(x,y) G_n(y)=-\gamma_n G_n(x)\,,	
\end{eqnarray}
where
\begin{eqnarray}
\gamma_n=\frac{1}{2}+2\sum\limits_{j=2}^{n+1}\frac{1}{j}-\frac{1}{(n+1)(n+2)}\,.	
\end{eqnarray}
and the eigen-functions
\begin{eqnarray}
G_n(x)\equiv x(1-x) C^{(3/2)}_n(2 x-1)\,,	
\end{eqnarray}
with $C^{(3/2)}_n(2 x-1)$ being order $3/2$ Gegenbauer polynomials. We can decompose the LCDA $\hat{\phi}^{(0)}_H(x)$ in the basis of $G_n$:
\begin{eqnarray}
\hat{\phi}^{(0)}_H(x)=\sum\limits_{n=0}^\infty \hat{\phi}_{H,n}^{(0)} G_n(x)\,,
\end{eqnarray}
where
\begin{eqnarray}
\hat{\phi}_{H,n}^{(0)}&=&\frac{4(2n+3)}{(n+1)(n+2)}\int_0^1 dx C_n^{(3/2)}(2x-1)\hat{\phi}^{(0)}_H(x)
\,.	
\end{eqnarray}
Hence, we can get
\begin{eqnarray}
\exp \left(\kappa C_{F} V \star\right) \hat{\phi}^{(0)}_H(x)=\sum\limits_{n=0}^\infty \hat\phi_{H,n}^{(0)}\left(\frac{\alpha_s(m_Q)}{\alpha_s(m_Z)}\right)^{-2\gamma_n C_F/\beta_0} G_n(x)\,.
\end{eqnarray}

Employing
\begin{eqnarray}
\int_0^1 dx\frac{1}{x}G_n(x)=(-1)^n\frac{1}{2}	\,,
\end{eqnarray}
and the decompositions
\begin{subequations}
\begin{eqnarray}
\delta(x-1/2)&=&\sum\limits_{n=0}^\infty  \frac{2(4n+3)}{(2n+1)(n+1)} (-1)^n \frac{(2n+1)!!}{(2n)!!}G_{2n}(x)\,,\\
-\frac{1}{2}\delta'(x-1/2)&=&\sum\limits_{n=0}^\infty  \frac{2(4n+5)}{(n+1)(2n+3)} (-1)^n \frac{(2n+3)!!}{(2n)!!}	G_{2n+1}(x)\,.
\end{eqnarray}
\end{subequations}

We get
\begin{eqnarray}\label{eq-kappa-c1}
\mathcal{K}_H^{\rm LL}&=&	\frac{1}{4}\sum\limits_{n=0}^\infty  \frac{2(4n+3)}{(2n+1)(n+1)} (-1)^n \frac{(2n+1)!!}{(2n)!!}\left(\frac{\alpha_s(m_Z)}{\alpha_s(m_Q)} \right)^{2 C_F\gamma_{2n}/\beta_0} \,,
\end{eqnarray}
for $H=\chi_{Q1}$, and
\begin{eqnarray}\label{eq-kappa-c0}
\mathcal{K}_H^{\rm LL}&=&	\frac{1}{4}\sum\limits_{n=0}^\infty  \frac{2(4n+5)}{(2n+3)(n+1)} (-1)^n \frac{(2n+3)!!}{(2n)!!}\left(\frac{\alpha_s(m_Z)}{\alpha_s(m_Q)} \right)^{2 C_F\gamma_{2n+1}/\beta_0} \,,
\end{eqnarray}
for $H=\chi_{Q0}\,,\chi_{Q2}$ and $h_Q$.

With the explicit expansion of the formal solution in Eq.~(\ref{eq:LLsolution}) in $\kappa$ and the formulas
\begin{subequations}
\begin{eqnarray}
&&\int_0^1 dy \frac{1}{y} V(y,x)=\frac{1}{x}\Bigg[\frac{3}{2}+\ln x\Bigg]\,,\\
&&\int_0^1 dz \int_0^1 d y \frac{1}{z} V(z,y) V(y,x) =\frac{1}{x}\Bigg[ \frac{9}{4}+\frac{3-2x}{1-x}\ln x+\ln^2 x+\mathrm{Li}_2(1-x)-\frac{\pi^2}{6}\Bigg]\,,
\end{eqnarray}
\end{subequations}
we can obtain the expansion of $\mathcal{K}_H^{\mathrm{LL}}$ in $\alpha_s$,
\begin{eqnarray}\label{eq-ll-2-chic1}
\mathcal{K}_H^{\rm LL}&=&1-\frac{\alpha_s(m_Z)}{4\pi} C_F\ln r^2 \Bigg[3-2\ln 2\Bigg]\nonumber\\
&&+\frac{\alpha_s^2(m_Z) }{(4\pi)^2}\ln^2r^2 \Bigg[ C_F^2\left(\frac{9}{2}-8\ln 2+\ln^2 2-\frac{\pi^2}{6}\right)+C_F\beta_0\left(\frac{3}{2}-\ln 2\right)\Bigg]+\mathcal{O}(\alpha_s^3)
\,,\nonumber\\
\end{eqnarray}
for $H=\chi_{Q1}$, and
\begin{eqnarray}\label{eq-ll-2-chic0}
\mathcal{K}_H^{\rm LL}&=&1-\frac{\alpha_s(m_Z)}{4\pi} C_F\ln r^2\Bigg[1-2\ln 2\Bigg]\nonumber\\
&&+\frac{\alpha_s^2(m_Z) }{(4\pi)^2}\ln^2 r^2 \Bigg[ C_F^2\left(-\frac{7}{2}+2\ln 2+\ln^2 2-\frac{\pi^2}{6}\right)+C_F\beta_0\left(\frac{1}{2}-\ln 2\right)\Bigg]+\mathcal{O}(\alpha_s^3) \,,\nonumber\\
\end{eqnarray}
for $H=\chi_{Q0}\,,\chi_{Q2}$ and $h_Q$.

\subsection{SDCs by combining the fixed-order results and the LL resummation}

In the following, we will improve the leading-twist SDCs by combining the NRQCD fixed-order results given in Sec.~\ref{sec-NRQCD} and the  LL resummation sketched in Sec.~\ref{subsec-LL} and Sec.~\ref{subsec-resum}.

To avoid double counting, it is necessary to subtract the terms of  $\mathcal{O}(\alpha_s^n \ln^n r)$
from the fixed-order SDCs.
Formally, we have
\begin{subequations}
\bqa\label{eq-NNLO-LL}
\mathcal{C}_{0,1}^{H,{\rm LO+LL}}&=&\mathcal{C}_{0,1}^{H,\rm LO}-
\mathcal{C}_{0,1}^{H,{\rm LL_0}}+\mathcal{C}_{0,1}^{H, \rm LL},\\
\mathcal{C}_{0,1}^{H,{\rm NLO+LL}}&=&\mathcal{C}_{0,1}^{H,\rm NLO}-
\mathcal{C}_{0,1}^{H,{\rm LL_1}}+\mathcal{C}_{0,1}^{H, \rm LL},\\
\mathcal{C}_{0,1}^{H,{\rm NNLO+LL}}&=&\mathcal{C}_{0,1}^{H,\rm NNLO}-
\mathcal{C}_{0,1}^{H,{\rm LL_2}}+\mathcal{C}_{0,1}^{H, \rm LL},
\eqa
\end{subequations}
where the superscripts `LO', `NLO' and `NNLO' denote the fixed-order SDCs accurate up to $\mathcal{O}(\alpha_s^0)$, $\mathcal{O}(\alpha_s^1)$ and $\mathcal{O}(\alpha_s^2)$ respectively,
and the superscripts `$\rm LL_0$', `$\rm LL_1$' and `$\rm LL_2$' signify the $\mathcal{C}_{0,1}^{H, \rm LL}$ truncated at
$\mathcal{O}(\alpha_s^0)$, $\mathcal{O}(\alpha_s^1)$ and $\mathcal{O}(\alpha_s^2)$ respectively, which have been explicitly
computed in Sec.~\ref{subsec-resum}.

In Tab.~\ref{tab-SDCs-squared}, we present the theoretical predictions on the squared leading-twist SDCs $|\mathcal{C}_{0, 1}^{H}|^2$ at various levels of accuracy.
In the table,  we also enumerate the values of $\mathcal{K}_H^{\rm LL}$, which are computed by applying the Eq.~(\ref{eq-kappa-c1})
and Eq.~(\ref{eq-kappa-c0})  with $\alpha_s(m_Z)=0.1181$, taken from the PDG, and $\alpha_s(m_c)=0.3240$ and $\alpha_s(m_b)=0.2151$,
 evaluated through the renormalization group running at two loop with the aid of  the package {\tt RunDec}.
 It is worth noting that, in order to accelerate the convergence, we have used the so-called Abel-Pad$\acute{\rm e}$ method~\cite{Bodwin:2016edd} to sum the series in Eq.~(\ref{eq-kappa-c1})  and Eq.~(\ref{eq-kappa-c0}).

\begin{table}[!htbp]
	\caption{Squared leading-twist SDCs $|\mathcal{C}_{0, 1}^{H}|^2$ at various levels of accuracy. We take $\mu_R=m_Z$ and $\mu_\Lambda=1$ GeV.
	}
	\label{tab-SDCs-squared}
	\setlength{\tabcolsep}{9pt}
	\renewcommand{\arraystretch}{1.3}
	\centering
	\begin{tabular}{|c|c|c|c|c|c|c|c|}
		\hline
		{$H$} & $\mathcal{K}_{H}^{\rm LL}$
		&LO
		& LO+LL
		&NLO
		&NLO+LL
		&NNLO
		&NNLO+LL\\
		\hline
		$\chi_{c0}$ & 0.859
		&7.96
		&5.87
		&8.01
		&6.47
		&8.85
		&7.90\\
		\hline
		$\chi_{c1}$ & 1.222
		&48.13
		&71.88
		&51.15
		&57.38
		&47.07
		&50.63\\
		\hline
		$\chi_{c2}$ & 0.859
		&16.04
		&11.86
		&10.89
		&8.37
		&8.87
		&7.54\\
		\hline
		$h_c$ & 0.859
		&24.00
		&17.73
		&17.31
		&13.41
		&13.51
		&11.50\\
		\hline
		$\chi_{b0}$ & 0.930
		&7.65
		&6.59
		&7.85
		&7.21
		&8.87
		&8.58\\
		\hline
		$\chi_{b1}$ & 1.145
		&49.08
		&64.25
		&47.72
		&50.31
		&45.52
		&46.78\\
		\hline
		$\chi_{b2}$ & 0.930
		&16.36
		&14.18
		&11.30
		&10.23
		&9.90
		&9.47\\
		\hline
		$h_b$ & 0.930
		&24.00
		&20.77
		&17.68
		&16.02
		&15.26
		&14.61\\
		\hline
	\end{tabular}
\end{table}

From Tab.~\ref{tab-SDCs-squared},  we have several observations. Firstly, we find the fixed-order predictions for charmonium production are
close to these for bottomonium production, however the LL resummation can give arise to some difference.
Secondly, we notice that the effect of the LL resummation can considerably change the LO results,
especially for charmonium production, for instance, it can change the LO results by more than $25\%$ for $\chi_{c0,2}$ and $h_c$ production, and
by around $50\%$ for $\chi_{c1}$ production. The magnitude of the LL resummation for bottomium production is roughly half of that for charmonium case.
Finally, it is worth noting that the effects of the LL resummation on the NLO and NNLO predictions are continuously becoming milder.
It can be explained by the fact that some of the LL resummation have been included in the radiative corrections. Concretely,
the $\mathcal{O}(\alpha_s\ln r)$ contribution has been included in the NLO prediction, while
both the $\mathcal{O}(\alpha_s\ln r)$ and $\mathcal{O}(\alpha^2_s\ln^2 r)$ contributions have been included in NNLO prediction.
As a consequence, the remaining contribution from the LL resummation can correspondingly reduce its effect on the NLO and NNLO predictions.

\section{phenomenology\label{sec-phen}}

To make concrete phenomenological predictions, we need to fix the various input parameters.
We take $m_Z=91.1876$ GeV from PDG, and the heavy quark pole masses $m_c=1.69$ GeV and $m_b=4.80$ GeV, as mentioned which are converted from their $\overline{\rm MS}$ masses.
We take the running QED coupling constant evaluate at the mass of $m_Z$ as $\alpha(m_Z)=1/128.943$~\cite{Sun:2016bel}.
We take the NRQCD factorization scale $\mu_\Lambda=1$ GeV.
The NRQCD LDMEs for $\chi_{QJ}$ and $h_{Q}$ are
approximated by the first derivative of the Schr\"{o}dinger radial wave function at origin through
\begin{subequations}
\begin{eqnarray}\label{eq-ldme-wave-funciton}
|\langle \mathcal{O} \rangle_{\chi_{cJ}}|^2&\approx&|\langle \mathcal{O} \rangle_{h_{c}}|^2 \approx \frac{3N_c}{2\pi} |R^\prime_{1P,c\bar{c}}(0)|^2
=\frac{3N_c}{2\pi}\times 0.075\, {\rm GeV^{5}},\\
|\langle \mathcal{O} \rangle_{\chi_{bJ}}|^2&\approx& |\langle \mathcal{O} \rangle_{h_{b}}|^2 \approx \frac{3N_c}{2\pi} |R^\prime_{1P,b\bar{b}}(0)|^2
=\frac{3N_c}{2\pi}\times 1.417\, {\rm GeV^{5}},
\end{eqnarray}
\end{subequations}
where the $1P$ radial wave functions at origin, evaluated in Buchm\"uller-Tye (BT) potential model, are taken from Ref.~\cite{Eichten:1995ch}.
In addition, we take $s_W^2=0.231$, and
the value of the total decay width of $Z$ boson $\Gamma_Z=2.4952$ GeV from
the PDG~\cite{ParticleDataGroup:2020ssz}.

With all ingredients in hand, we can compute the decay widths of $Z\to H+\gamma$.
The unpolarized decay widths at various levels of accuracy for charmonium production and bottomonium production
are separately tabulated in Tab.~\ref{tab-decay-rate-charm}  and Tab.~\ref{tab-decay-rate-bottom}.  In the tables,
we have included the uncertainties from ambiguity of renormalization scale and QCD higher-order corrections.
We should emphasize that the values of the Schr\"odinger wave functions may largely affect the theoretical predictions on the decay rates,
i.e., $|R^\prime_{1P,c\bar{c}}(0)|^2$ can range from 0.07 to 0.13 $\rm GeV^5$, $|R^\prime_{1P,b\bar{b}}(0)|^2$ can range from 0.93 to 2.07 $\rm GeV^5$ in Refs.~\cite{Eichten:1995ch,Chung:2021efj}, which
may change the central values of the decay rates of quarkonium production by roughly a factor of 2.

\begin{table}
	\caption{Unpolarized (total) decay widths for $Z\to {\rm charmonium}+\gamma$ at various levels of accuracy. The predictions are obtained by setting $\mu_R=m_Z/\sqrt{2}$.
The first error at accuracy of `NNLO' and `NNLO+LL' is estimated by varying $\mu_R$ from $m_Z/2$ to $m_Z$, while the second error is from the QCD higher-order corrections, which are estimated through $\alpha_s^3\sim 0.002$.}~\label{tab-decay-rate-charm}
	\setlength{\tabcolsep}{9pt}
	\renewcommand{\arraystretch}{1.3}
	\begin{center}
		\begin{tabular}{ccc ccc ccc ccc ccc ccc}
			\hline
			 \ \ \ Channel \ \ \
			&&&  \ \ \ Order \ \ \
			&&&  \ \ \ $\Gamma_{\rm total}(\rm eV)$\ \ \
			&&&\ \ \  $\rm Br(\times 10^{-9})$ \ \ \ \ \\
			\hline
			\ \ \multirow{4}{*}{$Z \to \chi_{c0}+\gamma$} \ \
			&&& LO
			&&& 0.939
			&&& $0.376$ \\
			&&& NLO
			&&& 0.946
			&&& $0.379$ \\
			&&& NNLO
			&&& $1.056^{+0.014+0.002}_{-0.012-0.002}$
			&&& $0.423^{+0.006+0.001}_{-0.005-0.001}$ \\
			&&& NNLO+LL
			&&& $0.944^{+0.013+0.002}_{-0.011-0.002}$
			&&& $0.374^{+0.005+0.001}_{-0.005-0.001}$ \\
			\hline
			\multirow{4}{*}{$Z \to \chi_{c1}+\gamma$}
			&&& LO
			&&& 5.687
			&&& 2.279\\
			&&& NLO
			&&& 6.066
			&&& 2.431  \\
			&&& NNLO
			&&& $5.529^{+0.033+0.011}_{-0.040-0.011}$
			&&& $2.216^{+0.013+0.004}_{-0.016-0.004}$ \\
			&&& NNLO+LL
			&&& $5.947^{+0.035+0.011}_{-0.042-0.011}$
			&&& $2.383^{+0.014+0.004}_{-0.017-0.004}$   \\
			\hline
			\multirow{4}{*}{$Z \to \chi_{c2}+\gamma$}
			&&& LO
			&&& 1.901
			&&& 0.762 \\
			&&& NLO
			&&& 1.259
			&&& 0.504 \\
			&&& NNLO
			&&& $0.997^{+0.053+0.002}_{-0.059-0.002}$
			&&& $0.399^{+0.021+0.001}_{-0.024-0.001}$ \\
			&&& NNLO+LL
			&&& $0.844^{+0.049+0.002}_{-0.054-0.002}$
			&&& $0.338^{+0.019+0.001}_{-0.022-0.001}$ \\
			\hline
			\multirow{4}{*}{$Z \to h_{c}+\gamma$}
			&&& LO
			&&& 19.231
			&&&  7.707 \\
			&&& NLO
			&&& 13.597
			&&&  5.449 \\
			&&& NNLO
			&&& $10.261^{+0.557+0.021}_{-0.625-0.021}$
			&&&  $4.112^{+0.223+0.008}_{-0.251-0.008}$ \\
			&&& NNLO+LL
			&&& $8.700^{+0.513+0.021}_{-0.575-0.021}$
			&&& $3.487^{+0.206+0.008}_{-0.230-0.008}$ \\
			\hline
		\end{tabular}
	\end{center}
	\label{Table-1}
\end{table}

\begin{table}
	\caption{Unpolarized (total) decay widths for $Z\to {\rm bottomonium}+\gamma$ at various levels of accuracy. The source of the theoretical uncertainties is the same as that in Tab~\ref{tab-decay-rate-charm}. ~\label{tab-decay-rate-bottom}}
	\setlength{\tabcolsep}{9pt}
	\renewcommand{\arraystretch}{1.3}
	\begin{center}
		\begin{tabular}{ccc ccc ccc ccc ccc ccc}
			\hline
			\ \ \ Channel \ \ \
			&&&  \ \ \ Order \ \ \
			&&&  \ \ \ $\Gamma_{\rm total}(\rm eV)$\ \ \
			&&&\ \ \  $\rm Br(\times 10^{-9})$ \ \ \ \ \\
			\hline
			\ \ \multirow{4}{*}{$Z \to \chi_{b0}+\gamma$} \ \
			&&& LO
			&&& 0.598
			&&& 0.240 \\
			&&& NLO
			&&& 0.615
			&&& 0.246 \\
			&&& NNLO
			&&& $0.704^{+0.012+0.001}_{+0.010-0.001}$
			&&& $0.282^{+0.005+0.001}_{-0.004-0.001}$ \\
			&&& NNLO+LL
			&&& $0.681^{+0.012+0.001}_{-0.010-0.001}$
			&&& $0.273^{+0.005+0.001}_{-0.004-0.001}$ \\
			\hline
			\multirow{4}{*}{$Z \to \chi_{b1}+\gamma$}
			&&& LO
			&&& 3.882
			&&& 1.556\\
			&&& NLO
			&&& 3.771
			&&& 1.511  \\
			&&& NNLO
			&&& $3.578^{+0.024+0.007}_{-0.028-0.007}$
			&&& $1.434^{+0.010+0.003}_{-0.011-0.003}$ \\
			&&& NNLO+LL
			&&& $3.676^{+0.025+0.007}_{-0.029-0.007}$
			&&& $1.473^{+0.010+0.003}_{-0.011-0.003}$   \\
			\hline
			\multirow{4}{*}{$Z \to \chi_{b2}+\gamma$}
			&&& LO
			&&& 1.323
			&&& 0.530 \\
			&&& NLO
			&&& 0.888
			&&& 0.356 \\
			&&& NNLO
			&&& $0.762^{+0.032+0.001}_{-0.035-0.001}$
			&&& $0.305^{+0.013+0.001}_{-0.014-0.001}$ \\
			&&& NNLO+LL
			&&& $0.729^{+0.031+0.001}_{-0.035-0.001}$
			&&& $0.292^{+0.012+0.001}_{-0.014-0.001}$ \\
			\hline
			\multirow{4}{*}{$Z \to h_{b}+\gamma$}
			&&& LO
			&&& 3.964
			&&& 1.589 \\
			&&& NLO
			&&& 2.860
			&&& 1.146 \\
			&&& NNLO
			&&& $2.419^{+0.093+0.005}_{-0.104-0.005}$
			&&& $0.969^{+0.037+0.002}_{-0.042-0.002}$ \\
			&&& NNLO+LL
			&&& $2.314^{+0.091+0.005}_{-0.102-0.005}$
			&&& $0.927^{+0.036+0.002}_{-0.041-0.002}$\\
			\hline
		\end{tabular}
	\end{center}
	\label{Table-1}
\end{table}

It is interesting to note that both the $\mathcal{O}(\alpha_s)$ and $\mathcal{O}(\alpha_s^2)$ corrections to $Z\to \chi_{Q2}/h_Q+\gamma$
are sizable and negative.
The LL resummation turns out to further decrease the decay widths.
Incorporating all the perturbative corrections and LL resummation reduces the LO prediction by roughly half of magnitude.
In contrast, both the  $\mathcal{O}(\alpha_s)$ and $\mathcal{O}(\alpha_s^2)$ corrections are moderate or even minor for other channels.
The situation is quite similar to the case in Ref.~\cite{Sang:2020fql}, where the radiative corrections
are significant for $e^+e^-\to \chi_{c2}+\gamma$  at B factory, however inconsiderable for $e^+e^-\to \chi_{c0,1}+\gamma$.

It is enlightening to compare the strengths of the decay widths for different quarkonium production.
For charmoinum production, we find that $h_c+\gamma$ production has the biggest branching fraction, followed
by $\chi_{c1}+\gamma$ production.  Although the branching fraction of $\chi_{c2}+\gamma$ is two times larger than that of
$\chi_{c0}+\gamma$ at LO,  their branching fractions are nearly the same at `NNLO+LL' accuracy.
For bottomonium production,  we notice that the branching fractions of $\chi_{b1}+\gamma$, $h_{b}+\gamma$,
$\chi_{c2}+\gamma$, $\chi_{b0}+\gamma$ make the most, second, third, and last strengths.

It is worth noting that, for the same quantum number of quarkonium, the branching fraction of charmonium production is larger than
that of bottomium production.
Finally, we estimate number of the quarkonium production at the proposed super $Z$ factories,
such as the $Z$-factory mode in CEPC, where $Z$ boson yield will reach $7\times 10^{11}$~\cite{CEPCStudyGroup:2018ghi}.
Thus it is expected that there will be several hundreds and thousands of charmonia and bottomonia production  through $Z\to H+\gamma$.
The signal for $Z\to \chi_{QJ}+\gamma$ production can be measured by probing $\chi_{QJ}$ with a recoiling hard photon, where $\chi_{QJ}$
can be reconstructed from their transition to $\gamma+J/\psi$ ($\gamma+\Upsilon$) with $J/\psi (\Upsilon) \to \ell \ell$.
Due to the low multiplicative branching ratio, and extra event-selection rules to suppress the backgrounds, it will be rather difficult to measure
  $Z\to \chi_{QJ}+\gamma$ in experiment.  Alternatively, the $\chi_{QJ}$ may be reconstructed through its hadronic decays, however,
 the experimental measurements on $\chi_{QJ}+\gamma$ are still challenging.  The condition for  $Z\to h_{Q}+\gamma$ is even worse. 

\section{summary~\label{sec-summary}}
In summary, we study the exclusive decay processes of $Z\to \chi_{QJ}/h_Q+\gamma$ in the NRQCD framework.
The amplitudes of all the helicity configurations and the unpolarized decay widths are evaluated up to $\mathcal{O}(\alpha_s^2)$.
It is the first time that the NRQCD factorization for $h_Q$ exclusive production at two loop is verified explicitly.
The LL of $m_Z^2/m_Q^2$ in the leading-twist SDCs are resummed to all orders of $\alpha_s$ by employing the LC factorization.
We find the radiative corrections are considerable  for $\chi_{Q2}$ and $h_Q$ productions, while are moderate or even minor for other channels.
We also notice that the LL resummation can change the LO results by more than 25\% for $\chi_{c0,2}$ and $h_c$ production, and
by around 50\% for $\chi_{c1}$ production.
However, effects of the LL resummation on the NLO and NNLO predictions are notably mitigated. We expect that several hundreds and thousands of charmonia and bottomonia will be produced  through $Z\to H+\gamma$ at the proposed super-Z factories.

\begin{acknowledgments}
The work of W.-L. S. and Y.-D. Z. is supported by the National Natural Science Foundation
of China under Grants No. 11975187.
The work of D.~Y. is supported in part by the National Natural Science Foundation of China under Grants No.~11635009.
\end{acknowledgments}

\section*{Appdendix}
\appendix
\section{Construction of helicity projectors\label{appendix-helicity-projectors}}

In this appendix, we present the helicity projectors $\mathcal{P}^{(H)}_{\lambda_1,\lambda_2}$, which have been used
to compute the helicity
amplitudes for $Z \to H(\lambda_1)+\gamma(\lambda_2)$ in Sec.~\ref{sec-NRQCD}.
We apply the similar technique applied in Refs.~\cite{Xu:2012uh,Zhang:2021ted}.

For the sake of convenience, we introduce an auxiliary transverse metric tensor and two auxiliary longitudinal vectors,
\begin{subequations}\label{eq-auxiliary}
	\begin{eqnarray}
	g_{\perp}^{\mu \nu}&=&g^{\mu \nu}+\frac{P^{\mu} P^{\nu}}{|\mathbf{P}|^{2}}-\frac{Q \cdot P}{m_{Z}^{2}|\mathbf{P}|^{2}}(P^{\mu} Q^{\nu}+Q^{ \mu} P^{\nu}),\\
	L_Z^\mu &=& \frac{1}{|\mathbf{P}|} \bigg(P^\mu-\frac{Q\cdot P}{m_Z^2}Q^\mu\bigg),\\
	L_{\chi_{Q J}}^\mu &=& \frac{1}{|\mathbf{P}|} \bigg(\frac{P\cdot Q}{m_Z m_{\chi_{Q J}}}P^\mu-\frac{m_{\chi_{Q J}}}{m_{Z}}Q^\mu\bigg),\\
	L_{h_{Q}}^{\mu}&=&\frac{1}{|\mathbf{P}|} \bigg(\frac{P\cdot Q}{m_Z m_{h_{Q}}}P^\mu-\frac{m_{h_{Q}}}{m_{Z}}Q^\mu\bigg),
	\end{eqnarray}
\end{subequations}
where $P$ and $Q$ denote the momenta of $H$ meson and $Z$ boson respectively.
It is obvious that the transverse metric tensor satisfies
\begin{subequations}
	\begin{eqnarray}
	&&g_{\perp \mu \nu} P^{\mu}=g_{\perp \mu \nu} Q^{\mu}=0,\\
	&&g_{\perp \mu}^{\mu}=2,\\
	&&g_{\perp \mu\alpha}g_\perp^{\alpha\nu}=g_{\perp \mu\alpha}g^{\alpha\nu}=g_{\perp \mu}^\nu,
	\end{eqnarray}
\end{subequations}
and the longitudinal vectors satisfy $L_{Z}^\mu Q_\mu=L_{\chi_{Q J}}^\mu P_\mu=L_{h_{Q}}^\mu P_\mu=0$.

We enumerate all the 8 helicity projectors
\begin{subequations}
	\begin{eqnarray}
	\mathcal{P}_{0,1}^{(\chi_{Q0})\mu \nu}&=&-\frac{1}{2}g_{\perp}^{\mu\nu},\\
	\mathcal{P}_{1,1}^{(\chi_{Q1})\mu \nu \alpha}&=&\frac{-1}{2m_{{Z}} |\mathbf{P}|} L_{Z}^{\mu}\epsilon^{\nu\alpha\rho \sigma}
	Q_{\rho} P_{\sigma},\\
	\mathcal{P}_{0,1}^{(\chi_{Q1})\mu \nu \alpha}&=&\frac{1}{2m_{Z} |\mathbf{P}|} L_{\chi_{c1}}^{\alpha}\epsilon^{\mu\nu\rho \sigma}
	Q_{\rho} P_{\sigma},\\
	\mathcal{P}_{2,1}^{(\chi_{Q2})\mu \nu\alpha\beta}&=&
      \frac{1}{4}(g_{\perp}^{\mu\nu}g_{\perp}^{\alpha\beta}-g_{\perp}^{\mu\alpha}g_{\perp}^{\nu\beta}
	-g_{\perp}^{\mu\beta}g_{\perp}^{\nu\alpha}),\\
	\mathcal{P}_{1,1}^{(\chi_{Q2})\mu \nu\alpha\beta}&=&
	\frac{-1}{2\sqrt{2}}L_{Z}^\mu(g_{\perp}^{\nu\alpha}L_{\chi_{Q2}}^\beta+g_{\perp}^{\nu\beta}L_{\chi_{Q2}}^\alpha),\\
	\mathcal{P}_{0,1}^{(\chi_{Q2})\mu \nu\alpha\beta}&=&
	\frac{-1}{2\sqrt{6}}g_{\perp}^{\mu\nu}(g_{\perp}^{\alpha\beta}+2L_{\chi_{Q2}}^\alpha L_{\chi_{Q2}}^\beta),\\
	\mathcal{P}_{0,1}^{(h_{Q})\mu \nu \alpha}&=&-\frac{1}{2}g_{\perp}^{\mu\nu} L_{h_{Q}}^{\alpha},\\
	\mathcal{P}_{1,1}^{(h_{Q})\mu \nu \alpha}&=&-\frac{1}{2}g_{\perp}^{\nu \alpha} L_{Z}^{\mu}.
	\end{eqnarray}
\end{subequations}

If we express the decay amplitudes of $Z \to H(\lambda_1)+\gamma(\lambda_2)$ as
\begin{subequations}
\begin{eqnarray}
\mathcal{A}^{(\chi_{Q0})}&=&\mathcal{A}_{\mu \nu}^{(\chi_{Q0})} \epsilon_{Z}^{\mu} \epsilon_{\gamma}^{* \nu},\\
\mathcal{A}^{(\chi_{Q1})}&=&\mathcal{A}_{\mu \nu \alpha}^{(\chi_{Q1})} \epsilon_{Z}^{\mu} \epsilon_{\gamma}^{* \nu}\epsilon_{\chi_{c1}}^{*\alpha},\\
\mathcal{A}^{(\chi_{Q2})}&=&\mathcal{A}_{\mu \nu \alpha\beta}^{(\chi_{Q2})} \epsilon_{Z}^{\mu} \epsilon_{\gamma}^{* \nu}\epsilon_{\chi_{Q2}}^{*\alpha\beta},\\
\mathcal{A}^{(h_{Q})}&=&\mathcal{A}_{\mu \nu \alpha}^{(h_{Q})} \epsilon_{Z}^{\mu}  \epsilon_{\gamma}^{* \nu}\epsilon_{h_{Q}}^{*\alpha},
\end{eqnarray}
\end{subequations}
where $\epsilon_{Z}$ and $\epsilon_{\gamma}$ represent the polarization vectors of the $Z$ boson and the photon respectively,
the helicity amplitude can be computed through
\begin{subequations}
\begin{eqnarray}
\mathcal{A}^{(\chi_{Q0})}_{0,1}&=&\mathcal{P}_{0,1}^{(\chi_{Q0})\mu \nu}\mathcal{A}_{\mu \nu}^{(\chi_{Q0})},\\
\mathcal{A}^{(\chi_{Q1})}_{0,1}&=&\mathcal{P}_{0,1}^{(\chi_{Q1})\mu \nu\alpha}\mathcal{A}_{\mu \nu\alpha}^{(\chi_{Q1})},\\
\mathcal{A}^{(\chi_{Q1})}_{1,1}&=&\mathcal{P}_{1,1}^{(\chi_{Q1})\mu \nu\alpha}\mathcal{A}_{\mu \nu\alpha}^{(\chi_{Q1})},\\
\mathcal{A}^{(\chi_{Q2})}_{2,1}&=&\mathcal{P}_{2,1}^{(\chi_{Q2})\mu \nu\alpha\beta}\mathcal{A}_{\mu \nu\alpha\beta}^{(\chi_{Q2})},\\
\mathcal{A}^{(\chi_{Q2})}_{1,1}&=&\mathcal{P}_{1,1}^{(\chi_{Q2})\mu \nu\alpha\beta}\mathcal{A}_{\mu \nu\alpha\beta}^{(\chi_{Q2})},\\
\mathcal{A}^{(\chi_{Q2})}_{0,1}&=&\mathcal{P}_{0,1}^{(\chi_{Q2})\mu \nu\alpha\beta}\mathcal{A}_{\mu \nu\alpha\beta}^{(\chi_{Q2})},\\
\mathcal{A}^{(h_{Q})}_{0,1}&=&\mathcal{P}_{0,1}^{(h_{Q})\mu \nu \alpha}\mathcal{A}_{\mu \nu \alpha}^{(h_{Q})},\\
\mathcal{A}^{(h_{Q})}_{1,1}&=&\mathcal{P}_{1,1}^{(h_{Q})\mu \nu \alpha}\mathcal{A}_{\mu \nu \alpha}^{(h_{Q})}.
\end{eqnarray}
\end{subequations}


\end{document}